\documentclass[prl,twocolumn]{revtex4}
\usepackage{epsfig}
\usepackage{bm}

\begin{document}

\title{Chaotic Hamiltonian dynamics of surface air temperature on daily to intraseasonal time scales}

\author{A. Bershadskii}

\affiliation{
ICAR, P.O. Box 31155, Jerusalem 91000, Israel
}

\begin{abstract}
  
  The surface air temperature daily records at the land-based locations with different climate conditions (from Arctic to Patagonia) have been studied on the daily to intraseasonal time scales (low frequency annual and seasonal variations have been removed by subtracting a wavelet regression from the daily records). It is shown that the power spectra of the daily time series exhibit a universal behaviour corresponding to the Hamiltonian distributed chaos. Global average temperature fluctuations (land-based data) and the tropical Pacific sea surface temperature fluctuations  (El Ni\~no/La Ni\~na phenomenon) have been also considered in this context. It is shown that the practical smooth predictability for the surface air temperature dynamics is possible at least up to the fundamental (pumping) period of the distributed chaos. 

\end{abstract}

\maketitle

\section{Introduction}

  The digital data-bases with numerous long-term local meteorological records open new possibilities for a comparative analysis of the meteorological data over areas with different climate conditions \cite{bar}. This analysis is crucial for understanding of the differences
and universal properties in the dynamics of the meteorological processes. The (near) surface temperature is one of the most significant meteorological characteristics. Due to dependence on many different factors (such as latitude, elevation, atmospheric and oceanic circulations, coastal or interior location, terrain, etc.) the land-based (near) surface air temperature is not expected to posses universal properties, especially on the daily to intraseasonal time scales. On the other hand, chaotic nature of the temperature dynamics at these time scales can result in some universality (see, for instance, Ref. \cite{sp}). In present note we will try to find out these universal features, which appear to be rather strong. Of course, a comprehensive analysis of this universality on global scales is far beyond of our present scope. But already a preliminary glance at some geographical locations scattered around the globe reveal a rather interesting picture, that can justify a more comprehensive analysis. These locations were chosen mostly because of existence of sufficiently long and continuous daily records of the (near) surface temperature and of a minimization of the urban effect.\\

  There is another approach: a large-scale space average of the local temperature fluctuations. This approach can provide a global average temperature daily time series (see, for instance, Ref. \cite{ber1},\cite{ber2} and references therein). An analysis of these time series can be then compared with the analysis of the space localized data. \\
  
  Daily data for the surface air temperature over oceans as well as for the sea surface temperature - SST (i.e. the water temperature near the ocean's surface) are, naturally, much less available for a sufficiently long period of time. However, for the most interesting area of the Pacific ocean (its tropical part) the data have been accumulated in relation to the important El Ni\~no/La Ni\~na phenomenon. This phenomenon affects the surface air temperature across the globe and it is interesting to compare its chaotic properties with those of the surface air temperature over land.\\

\section{Distributed chaos in the Hamiltonian dynamical systems}

 A Hamiltonian dynamical system can be described by a Hamiltonian scalar function - $H(q,p,t)$ of the two vectors - position q and momentum p, and of the time t. The variables are related by the evolution equation 
$$
\frac{dq}{dt}=\frac{\partial H}{\partial p}, ~~~~~~~~~~\frac{dp}{dt}=-\frac  {\partial H}{\partial q}  \eqno{(1)}
$$

Let us expand analytically at frequency $f=0$ the power spectrum  of the position $q(t)$ 
$$
E_q(f) = I_q + I_q^{(1)}f +I_q^{(2)}f^2 +......   \eqno{(2)}
$$
and the power spectrum of the momentum $p(t)$ 
$$
E_p(f) = I_p + I_p^{(1)}f +I_p^{(2)}f^2 +......   \eqno{(2)}
$$
For a time independent Hamiltonian or for the situations when the action $I$ is an adiabatic invariant \cite{suz}
$$
I_p = c\cdot I \eqno{(3)}
$$
where $c$ is a dimensionless constant. Therefore $I_p$ can be considered as a momentum-action and, analogously, $I_q$ can be considered as a position-action. 

  A relationship between characteristic velocity $v_c$ and the characteristic frequency $f_c$ can be found using the dimensional considerations \cite{suz}:
$$    
v_c \propto I^{1/2} f_c^{1/2}  \eqno{(4)}
$$
or (using Eq. (3))
$$
v_c \propto I_p^{1/2} f_c^{1/2}  \eqno{(5)}
$$
  Analogously, depending on initial and boundary conditions,
$$  
v_c   \propto I_q^{1/2} f_c^{3/2}  \eqno{(6)}
$$
In a more general form
$$
v_c \propto f_c^{\alpha}  \eqno{(7)}
$$
where $\alpha$ can take the two values $\alpha =1/2$ or $\alpha = 3/2$. \\

   For most of the smooth chaotic systems decaying part of the frequency spectrum has an exponential form 
$$
 E(f) \propto \exp-(f/f_c) \eqno{(8)},
$$    
while for smooth chaotic Hamiltonian systems it often has a stretched exponential form \cite{b1}
$$
E(f ) \propto \int_0^{\infty} P(f_c) \exp -(f/f_c)~ df_c  \propto \exp-(f/f_0)^{\beta}  \eqno{(9)}
$$
 where  $P(f_c)$ is a weight (distribution) of the exponentials with the characteristic frequency $f_c$. 
\begin{figure} \vspace{-1.2cm}\centering
\epsfig{width=.38\textwidth,file=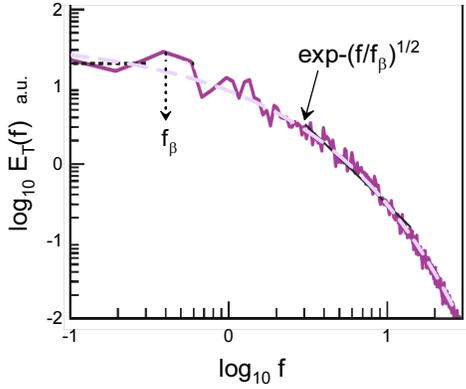} \vspace{-2.6cm}
\caption{Power spectrum of the temperature fluctuations measured using real-space probes in a direct numerical simulation of the Rayleigh-B\'{e}nard (thermal) convection at centre of a cubical box. }
\end{figure}
\begin{figure} \vspace{-1.473cm}\centering
\epsfig{width=.45\textwidth,file=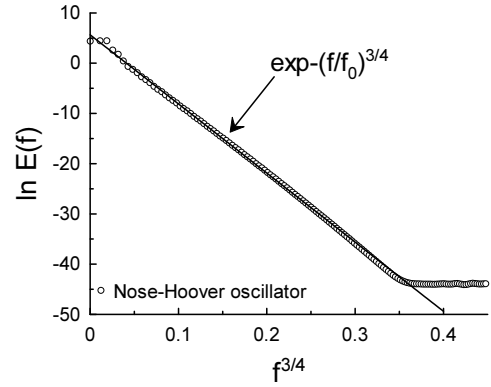} \vspace{-4.07cm}
\caption{Power spectrum for the Nose-Hoover oscillator. The data for the computations were taken from Ref. \cite{gen}). }
\end{figure}
\begin{figure} \vspace{-0.2cm}\centering
\epsfig{width=.45\textwidth,file=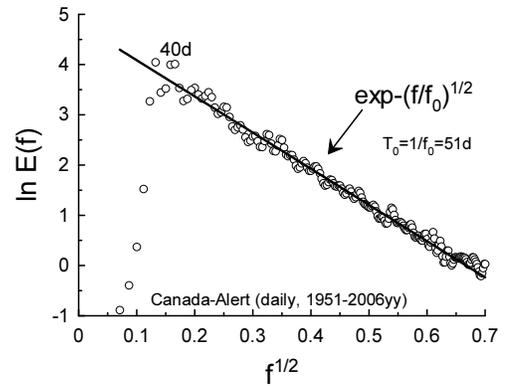} \vspace{-4.2cm}
\caption{The power spectrum of the daily mean surface air temperature measured for the period 1951-2006yy at Alert location (Nunavut, Canada).}
\end{figure}

If one denotes the distribution of the $v_c$ as $\mathcal{P} (v_c)$, then 
$$
\mathcal{P} (v_c) dv_c \propto P(f_c) df_c  \eqno{(10)}
$$
or, taking into account the Eq. (7),
$$
P(f_c)  \propto f_c^{\alpha -1} ~\mathcal{P} (v_c(f_c)) \eqno{(11)}
$$ 
  On the other hand, for the stretched exponential spectrum  Eq. (9) the $P(f_c)$ asymptote at $f_c \rightarrow \infty$ is 
$$
P(f_c) \propto f_c^{-1 + \beta/[2(1-\beta)]}~\exp(-bf_c^{\beta/(1-\beta)}) \eqno{(12)}
$$
with a constant $b$ \cite{jon}.

  From the Eqs. (7),(11) and (12) one can conclude that if the distribution $\mathcal{P} (v_c)$ is Gaussian (with zero mean), then the $\beta$ is related to the $\alpha$ by the equation
$$
\beta = \frac{2\alpha}{1+2\alpha}   \eqno{(13)}
$$

  It follows from the Eq. (13) that for $\alpha = 3/2$ (the Eq. (6)) the $\beta =3/4$, and for $\alpha = 1/2$ (the Eq. (5)) the $\beta =1/2$, depending on the initial and boundary conditions.\\
  
    Figure 1 shows, as an example, a power spectrum for the temperature fluctuations measured using real-space probes in a direct numerical simulation (DNS) of the Rayleigh-B\'{e}nard (thermal) convection at centre of a cubical box (the spectral data were taken from the Ref. \cite{kv}). In this DNS the dynamical equations under Boussinesq approximation were numerically solved with conducting boundary conditions for the temperature field at the horizontal walls (heating from below and cooling from above) and insulating boundary conditions at the side  walls. For the velocity field the no-slip boundary conditions were used at all the walls. The DNS was performed for the Prandtl number $Pr = 1$ and Rayleigh number $Ra = 10^8$. Absence of mean velocity at the centre of the cube allows avoid complications related to the Taylor's hypothesis \cite{kv}. The dashed (blue) curve in the Fig. 1 indicates correspondence to the stretched exponential Eq. (9) with $\beta = 1/2$. The vertical dotted arrow indicates position of the frequency $f_{\beta}$. One can see that the entire distributed chaos is tuned to the low-frequency coherent structures in this case. The Hamiltonian models of the Rayleigh-B\'{e}nard convection can be found in the Refs. \cite{tg},\cite{gl}.  \\
    
    Another relevant example is shown in Fig. 2. It is a power spectrum for the $x(t)$ variable of the Nose-Hoover oscillator. This oscillator can be considered as a harmonic oscillator contacting with a thermal bath. It is a Hamiltonian system \cite{spot}. The data used for computation of the spectrum were taken from the site \cite{gen}. The maximum entropy method with an optimal resolution \cite{oh} has been used for the computation. The straight line indicates (in the appropriately chosen scales) correspondence to the Eq. (9) with $\beta = 3/4$. \\

\section{Northern Hemisphere}  
\begin{figure} \vspace{-1.4cm}\centering
\epsfig{width=.45\textwidth,file=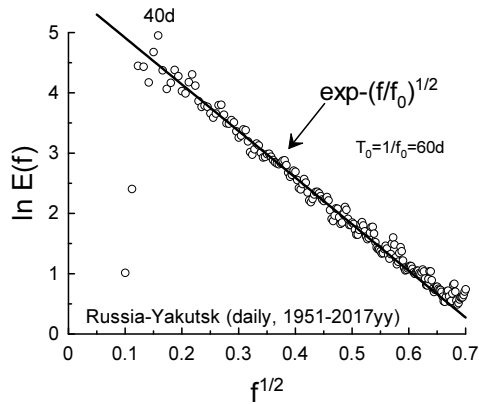} \vspace{-3.76cm}
\caption{As in Fig. 3 but for Yakutsk (Russia, Siberia).}
\end{figure}
\begin{figure} \vspace{-1.5cm}\centering
\epsfig{width=.45\textwidth,file=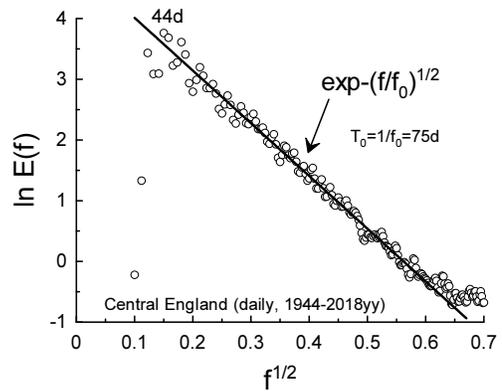} \vspace{-3.7cm}
\caption{As in Fig. 3 but for Central England for the period 1944-2018yy.}
\end{figure}

 Due to Hamiltonian nature of most of the climate and meteorological models \cite{gl},\cite{she}-\cite{gl2} one can expect that power spectra of the local surface air temperature fluctuations will exhibit the stretched exponential decay Eq. (9) with the $\beta =1/2$ or the $\beta =3/4$. The power spectra of the local surface air temperature fluctuations at 11 different geographical locations around the globe are shown in Figs. 3-14. The spectra have been computed using the maximum entropy method with an optimal resolution \cite{oh}. Low frequency annual and seasonal variations have been removed by subtracting a wavelet (symmlet of order four) regression from the daily time series \cite{ogd}. This subtraction removes also the long-term trends and the remaining time series represent the daily to intraseasonal dynamics only (cf. Refs. \cite{wh},\cite{ven2}).\\
  
 Let us start from  the Alert ( Qikiqtaaluk Region, Nunavut, Canada) - the northernmost permanently inhabited location on the globe, far north of the Arctic Circle (Latitude:	82.5018 and Longitude: -62.3481, elevation: 30 m). The climate is of a very dry polar type. Figure 3 shows power spectrum of the daily mean surface air temperature measured for the period 1951-2006yy (the data were taken from the site \cite{can}). The straight line in the figure (the best fit) indicates, in the appropriately chosen scales, the stretched exponential decay Eq. (9) with the $\beta =1/2$ (the Hamiltonian distributed chaos, see previous Section). The fundamental (pumping) period $T_f \simeq 40$d. \\ 
 
  Yakutsk (Russia, Sakha Republic, Siberia) - located 450 kilometers south of the Arctic Circle (Latitude: 62.0354 and Longitude: 129.6755, elevation: 126  m). Yakutsk has an extreme subarctic climate with a distinct inland location. The winters are extremely long and cold (the lowest temperature recorded here -64.4 $^0$C), while summers are warm (daily max. temperatures can exceed +30 °$^0$C). The climate is rather dry. Figure 4 shows power spectrum of the daily mean surface air temperature measured for the period 1951-2017yy (the data were taken from the site \cite{asi}). The straight line in the figure (the best fit) indicates, in the appropriately chosen scales, the stretched exponential decay Eq. (9) with the $\beta =1/2$. The fundamental (pumping) period $T_f \simeq 40$d. \\

  Next, let us look at the Central England with its temperate maritime climate. Although the urban effect is rather considerable at this region (a triangular area enclosed by London, Bristol, and Lancashire - Latitude: 53.7632, Longitude: -2.7044) there is a long continuous data set of the mean daily temperatures \cite{hadcet},\cite{plf} that allows to compare the temperature power spectra for a modern (1944-2018yy) and for an historical (1858-1943yy) periods (see Figs. 5 and 6). An again the both decaying parts of the spectra correspond to the stretched exponential Eq. (9) representing the Hamiltonian distributed chaos with the same $\beta =1/2$. The fundamental (pumping) period in both these cases: $T_f \simeq 44$d, is also close to that observed in the previous cases.\\
\begin{figure} \vspace{-1.5cm}\centering
\epsfig{width=.45\textwidth,file=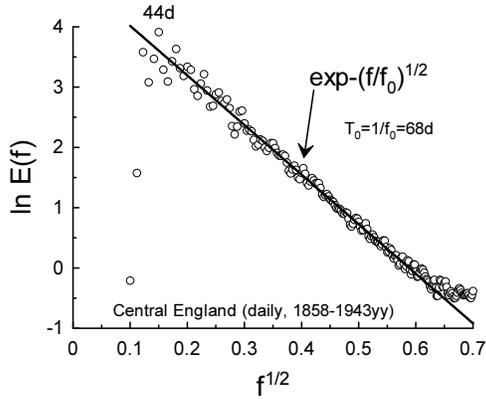} \vspace{-4.3cm}
\caption{As in Fig. 5 but for the Historical period 1858-1943yy.}
\end{figure}
\begin{figure} \vspace{-1.8cm}\centering
\epsfig{width=.45\textwidth,file=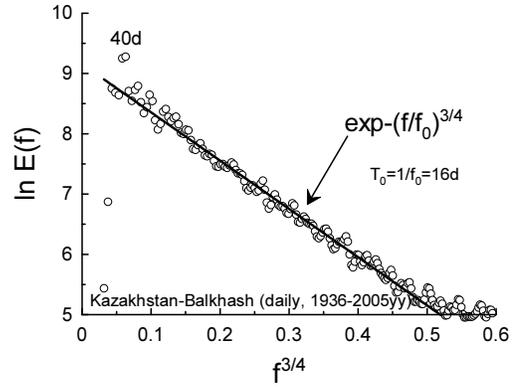} \vspace{-4cm}
\caption{As in Fig. 3 but for Balkhash (Kazakhstan, Central Asia).}
\end{figure}
\begin{figure} \vspace{-0.5cm}\centering
\epsfig{width=.45\textwidth,file=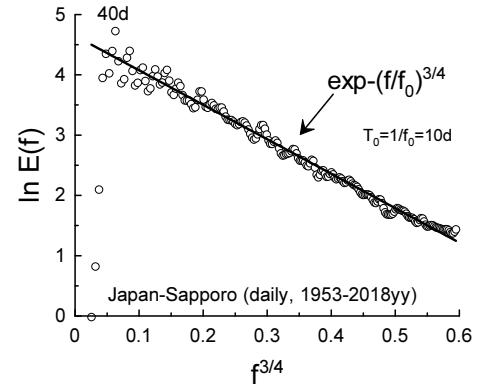} \vspace{-4.4cm}
\caption{As in Fig. 3 but for Sapporo (Japan).}
\end{figure}

\begin{figure} \vspace{-0.4cm}\centering
\epsfig{width=.45\textwidth,file=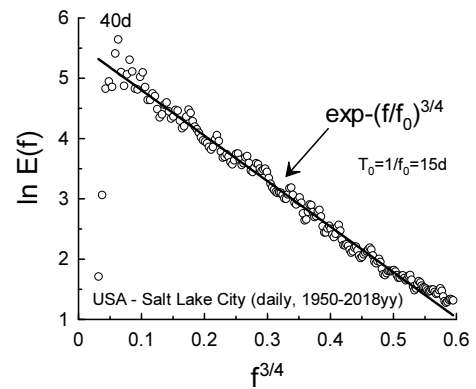} \vspace{-4.1cm}
\caption{As in Fig. 3 but for Salt Lake City (Utah, USA).}
\end{figure}

  Next is the Balkhash city (Kazakhstan, Central Asia - Latitude: 46.8435, Longitude: 74.9810, elevation: 440 m) with its semi-arid (continental) climate. Although the city is located on the northern shore of the Lake Balkhash (one of the largest lakes in Asia) precipitation is low. Figure 7 shows power spectrum of the daily mean surface air temperature measured for the period 1936-2005yy (the data were taken from the site \cite{eu}). The straight line in the figure (the best fit) indicates, in the appropriately chosen scales, the stretched exponential decay Eq. (9) now with the $\beta =3/4$ (also the Hamiltonian distributed chaos, see previous Section). The fundamental (pumping) period $T_f \simeq 40$d.\\
  
  Sapporo (Japan, Latitude: 43.0625,  Longitude: 141.3536, elevation: 26 m) has a humid continental climate  with a significant rainfall. Figure 8 shows power spectrum of the daily mean surface air temperature measured for the period 1953-2018yy (the data were taken from the site \cite{noaa}). The straight line in the figure (the best fit) indicates, in the appropriately chosen scales, the stretched exponential decay Eq. (9) with the $\beta =3/4$ (cf. Fig. 7). The fundamental (pumping) period $T_f \simeq 40$d.\\
  
  As the Sapporo the Salt Lake City (Utah, USA - Latitude: 40.7608, Longitude: -111.8910, elevation 1304 m) has a hot-summer humid continental climate. The Pacific Ocean and the Great Salt Lake are considered as the primary influences on the weather of this location. Figure 9 shows power spectrum of the daily mean surface air temperature measured for the period 1950-2018yy (the data were taken from the site \cite{noaa}). The straight line in the figure (the best fit) indicates the stretched exponential decay Eq. (9) with the $\beta =3/4$ (cf. Figs. 7 and 8). The fundamental (pumping) period $T_f \simeq 40$d.\\
  
\section{Tropics}
\begin{figure} \vspace{-0.4cm}\centering
\epsfig{width=.45\textwidth,file=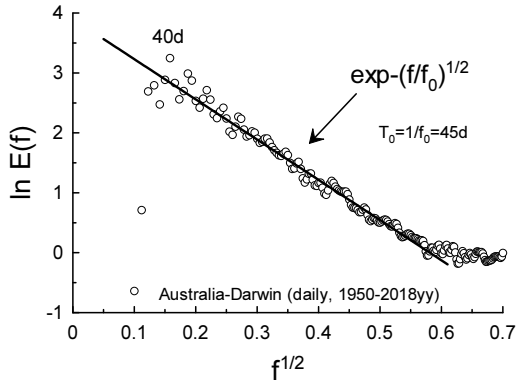} \vspace{-4.4cm}
\caption{As in Fig. 3 but for Darwin (Northern Australia).}
\end{figure}
\begin{figure} \vspace{-1.25cm}\centering
\epsfig{width=.45\textwidth,file=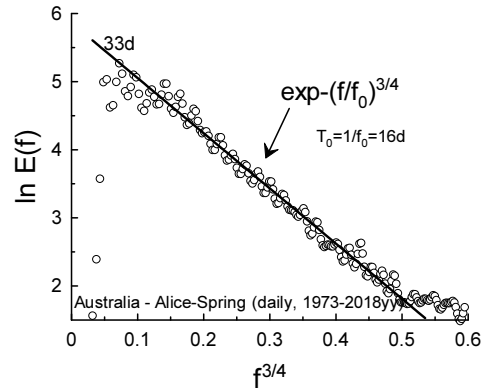} \vspace{-3.54cm}
\caption{As in Fig. 3 but for Alice Spring (Central Australia).}
\end{figure}
\begin{figure} \vspace{-0.5cm}\centering
\epsfig{width=.45\textwidth,file=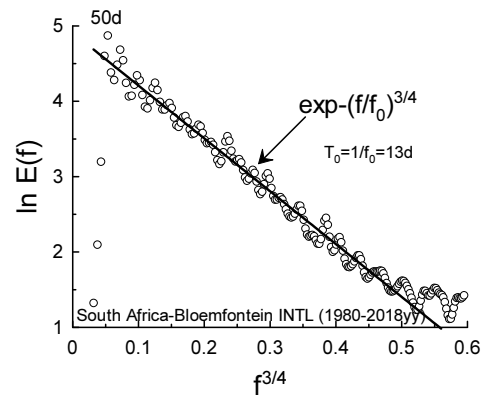} \vspace{-4.35cm}
\caption{As in Fig. 3 but for Bloemfontein (central South Africa).}
\end{figure}
   Now let us jump to the tropics - Darwin (in the top end of the Northern Australia, Latitude: -12.4634, Longitude: 130.8456, elevation: 37 m). It has a tropical savanna climate, that is characterized by very stable mean daily temperature all year round, with clear distinct dry and wet seasons. The wet season is characterised by monsoonal rains and high humidity, while the dry season is characterised by cool nights and sunny warm days. Figure 10 shows power spectrum of the daily mean surface air temperature measured for the period 1950-2018yy (the data were taken from the site \cite{noaa}). The straight line in the figure (the best fit) indicates the stretched exponential decay Eq. (9) with the $\beta =1/2$. The fundamental (pumping) period $T_f \simeq 40$d. The power spectrum is rather similar to those observed at the high latitudes (cf. Figs. 3-6). \\
   
\section{Southern Hemisphere}

   Dissimilarity between the Northern and Southern hemispheres geographic setting in respect of the land-ocean distribution could result in the significant differences in the time variability of the surface air temperature. However, as it will be seen from the figures 11-14, the main properties of the Hamiltonian distributed chaos at the daily to intraseasonal time scales are preserved also for the Southern Hemisphere. Only values of the fundamental (pumping) period are different. \\
   
   The Alice Spring (central Australia, Latitude: -23.6980, Longitude: 133.8807, elevation: 545 m) has a desert (a semi-arid) climate with little rainfall throughout the year, hot summers and cold winters. Figure 11 shows power spectrum of the daily mean surface air temperature measured for the period 1973-2018yy (the data were taken from the site \cite{noaa}). The straight line in the figure (the best fit) indicates the stretched exponential decay Eq. (9) with the $\beta =3/4$. The fundamental (pumping) period $T_f \simeq 33$d (cf. Figs. 7-9).\\
   
   The Bloemfontein (central South Africa, Latitude: -29.0852, Longitude: 26.1596, elevation: 1395 m) has a semi-arid climate with hot summer and dry, cooler winters. Figure 12 shows power spectrum of the daily mean surface air temperature measured for the period 1980-2018yy (the data were taken from the site \cite{noaa}). The straight line in the figure (the best fit) indicates the stretched exponential decay Eq. (9) with the $\beta =3/4$. The fundamental (pumping) period $T_f \simeq 50$d, that can be a consequence of the Madden-Julian Oscillation \cite{wh},\cite{ven2},\cite{b3}.\\
   
   The Santa Rosa (central Argentina, Latitude: -36.6209, Longitude: -64.2912369, elevation: 175 m) has a humid subtropical climate with hot (warm) summers and dry, chilly winters. Figure 13 shows power spectrum of the daily mean surface air temperature measured for the period 1973-2018yy (the data were taken from the site \cite{noaa}). The straight line in the figure (the best fit) indicates the stretched exponential decay Eq. (9) with the $\beta =3/4$. The fundamental (pumping) period $T_f \simeq 31$d (cf. Figs. 11 and 12).\\
\begin{figure} \vspace{-1.5cm}\centering
\epsfig{width=.45\textwidth,file=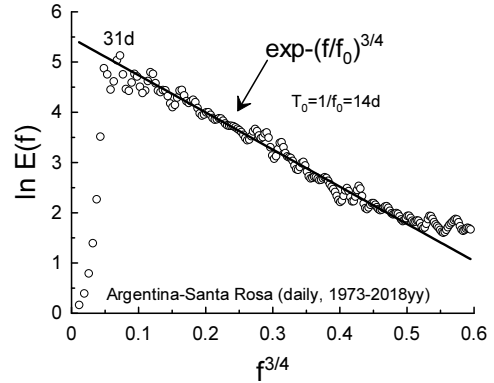} \vspace{-4.23cm}
\caption{As in Fig. 3 but for Santa Rosa (central Argentina).}
\end{figure}

\begin{figure} \vspace{-0.5cm}\centering
\epsfig{width=.45\textwidth,file=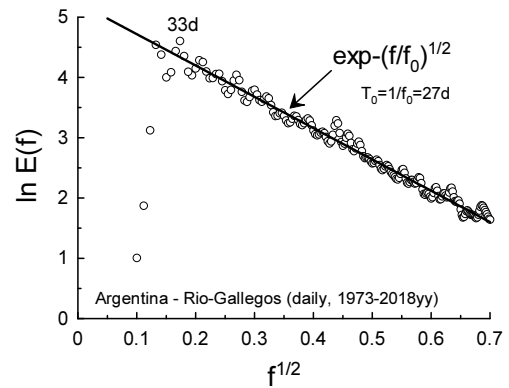} \vspace{-4cm}
\caption{As in Fig. 3 but for Rio Gallegos (Patagonia, Argentina).}
\end{figure}
 
   The Rio Gallegos (Patagonia, Argentina, Latitude: -51.6230, Longitude: -69.2168, elevation: 20 m) has a cold semi-arid climate (with a low precipitation and very strong winds).  Figure 14 shows power spectrum of the daily mean surface air temperature measured for the period 1973-2018yy (the data were taken from the site \cite{noaa}). The straight line in the figure indicates the stretched exponential decay Eq. (9) with the $\beta =1/2$. The fundamental (pumping) period $T_f \simeq 33$d.

\section{Global temperature}

  The apparent universality of the space localized spectra on the daily to intraseasonal time scales suggests a possibility that the global (average) surface temperature fluctuations (see Introduction) can be dominated by the Hamiltonian distributed chaos as well.

  Figure 15 shows a global average surface temperature anomaly (relative to the Jan. 1951 - Dec. 1980 average). The land-based daily data for the period 1936-2018yy were taken from the site \cite{ber2}. The entire daily time series shown in the Fig. 15 is statistically non-stationary and a certain detrending is necessary before a spectral analysis. But before the detrending it should be noted that the time sub-series for the period 1936-1976yy is approximately statistically stationary, and the spectral computation can be performed for this sub-series without detrending. Figure 16 shows power spectrum corresponding to this statistically stationary sub-set of the data.  The straight line in the figure (the best fit) indicates the stretched exponential decay Eq. (9) with the $\beta =1/2$.
\begin{figure} \vspace{-1cm}\centering
\epsfig{width=.45\textwidth,file=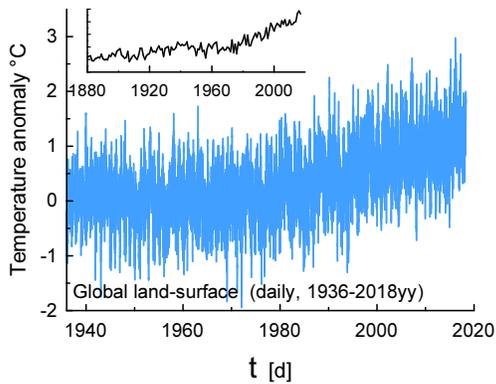} \vspace{-4.5cm}
\caption{Global average surface temperature anomaly for 1936-2018yy. The land-based daily data were taken from the Ref. \cite{ber2}. The insert shows annual mean of the surface temperature anomaly (land-based) for 1880-2017yy.} 
\end{figure}
\begin{figure} \vspace{-0.6cm}\centering
\epsfig{width=.45\textwidth,file=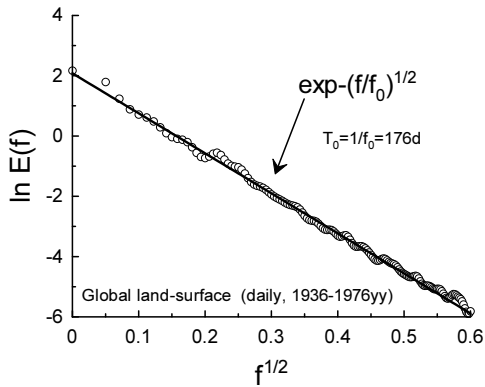} \vspace{-3.94cm}
\caption{Power spectrum corresponding to the sub-set (1936-1976yy) of the daily data shown in the Fig. 15.} 
\end{figure}
\begin{figure} \vspace{-1.7cm}\centering
\epsfig{width=.45\textwidth,file=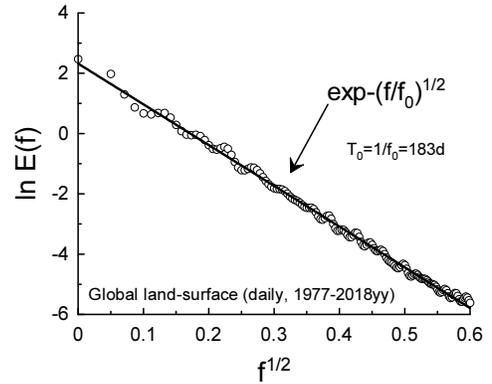} \vspace{-3.77cm}
\caption{Power spectrum corresponding to the sub-set (1977-2018yy) of the daily data shown in the Fig. 15.} 
\end{figure} 
   
\begin{figure} \vspace{+0.2cm}\centering
\epsfig{width=.45\textwidth,file=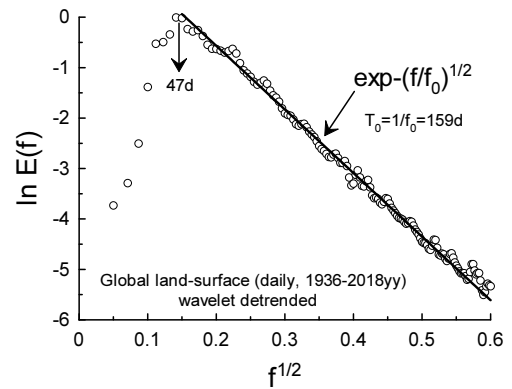} \vspace{-4.05cm}
\caption{Power spectrum corresponding to the wavelet regression detrended daily time series shown in the Fig. 15. } 
\end{figure}
     The second part of the time series, for the time period 1977-2018yy, can be detrended using the simplest linear regression method. Figure 17 shows corresponding power spectrum. The straight line in the figure (the best fit) indicates the stretched exponential decay Eq. (9) with the $\beta =1/2$.
     
      On can conclude that the daily global average temperature fluctuations are dominated by the Hamiltonian distributed chaos with $\beta =1/2$. 
      
      It is interesting that $T_0$ period is nearly a half-year in this case (see Figs. 16 and 17). The global temperature fluctuations can be sensitive to this period due to  dissimilarity between the Northern and Southern hemispheres geographic setting in respect of the land-ocean distribution. There is also more of a difference in solar radiation between winter and summer in the Southern hemisphere than in the Northern due to the Earth orbit eccentricity. The northward cross-equatorial ocean heat transport plays its role as well \cite{kan}.

   Now let us make a detrending removing low-frequency annual and seasonal variations by subtracting a wavelet regression from the daily time series shown in the Fig. 15. The symmlet wavelets were used for this purpose for the local temperature fluctuations in the previous Sections. They are a modified version of Daubechies wavelets (with increased symmetry) \cite{ogd}.  A special (and simplest) case of the Daubechies wavelet is the Haar wavelet. It is especially effective for analysis of signals with abrupt fluctuations, that is the case of the daily global temperature fluctuations. Therefore this wavelet regression has been used for the daily time series shown in the Fig. 15. Figure 18 shows power spectrum computed for the Haar wavelet detrended daily time series (1936-2018yy).

The straight line in the figure (the best fit) indicates the stretched exponential decay Eq. (9) with the $\beta =1/2$. The fundamental (pumping) period $T_f \simeq 47$d.

\section{Tropical Pacific SST}
\begin{figure} \vspace{-0cm}\centering
\epsfig{width=.5\textwidth,file=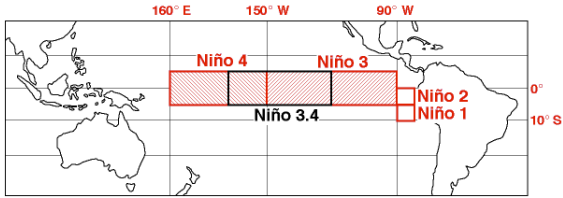} \vspace{-9.5cm}
\caption{Map of the tropical Pacific regions.}
\end{figure}
\begin{figure} \vspace{-0.1cm}\centering
\epsfig{width=.45\textwidth,file=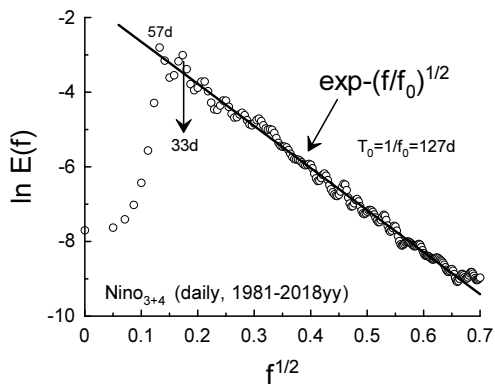} \vspace{-4.2cm}
\caption{Power spectrum corresponding to the wavelet regression detrended  daily index Ni\~n$\rm{o}_{3+4}$ for 1981-2018yy period. } 
\end{figure}
 It is well known that the El Ni\~no/La Ni\~na (or ENSO: El Ni\~no - Southern Oscillation) phenomenon causes major changes in the global atmospheric circulation and, as a consequence, changes of  the surface air temperatures across the globe (see, for instance, Refs. \cite{hr}-\cite{fan} and references therein). Therefore it is interesting to look at the chaotic properties of this phenomenon on the daily to intraseasonal time scales and compare these properties with those of the surface air temperatures over land. The important point here is that the El Ni\~no/La Ni\~na indices: Ni\~no1,...,Ni\~no4, are the area averaged SST ({\it sea} surface temperature, i.e. the water temperature near the ocean's surface). The Ni\~no3 (5N-5S, 150W-90W region) and the Ni\~no4 (5N-5S, 160E-150W region) indices together cover a rather wide and representative area in the tropical Pacific (see Fig. 19). In this area the El Ni\~no/La Ni\~na phenomenon is mainly observed. Figure 20 shows power spectrum corresponding to the weighted index Ni\~n$\rm{o}_{3+4}$, computed from the indices Ni\~no3 and Ni\~no4 for the entire region (5N-5S, 160E-90W), for 1981-2018yy period. A simple weighting has been used to take into account the difference between the areas corresponding to the Nino3 and Nino4 indices. The data for computations were taken from the site \cite{nl} and the low frequency annual and seasonal variations have been removed by subtracting the wavelet (symmlet of order four, used above for the local surface air temperature fluctuations) regression from the daily time series. The straight line in the figure (the best fit) indicates the stretched exponential decay Eq. (9) with the $\beta =1/2$.
   
\section{Predictability}

  After seminal paper Ref. \cite{l2} the predictability problem obtained several new trends which dominate its development even now. 
  
1.  There were introduced two main types of predictability. Let us define them as the {\it smooth} and {\it rough} ones \cite{b4}. At the {\it smooth} predictability  the predictability horizon can be indefinitely extended by reducing the initial error. At the {\it rough} predictability the predictability horizon cannot be indefinitely extended by reducing the initial error. It was generally assumed that the real weather forecasts belong to the second type of the predictability with the finite horizon which cannot be larger than two weeks.  

2. Following to the Kolmogorov's scaling paradigm \cite{my} it was assumed that the power-law spectra completely determine the real weather forecasts. Since this assumption the power-law spectra were the main theoretical tool in this respect and  there were many attempts to find such spectra for the simulations and naturally observed data (see, for instance the recent Ref. \cite{sz} and references therein). 

3. It was assumed that  that for the power-law space (wavenumber) spectra of the kinetic energy $E(k) \propto k^{-\gamma}$ with the $\gamma < 3$ the predictability is rough. This assumption was extended on the case $\gamma =3$ in recent Ref. \cite{leu}.

4. The main predictability efforts were concentrated on the space properties of the velocity field and the temperature field was considered as a secondary one for the weather predictability problem \cite{v}.  \\

  It was shown in the recent Ref. \cite{b4} that the stretched exponential frequency spectra Eq. (9) can be considered as a strong indicator of the smooth predictability. The first example given in present paper - the Rayleigh-B\'{e}nard thermal convection (see Section II), provides a clue to the situation. The entire frequency range covered by the distributed chaos with the stretched exponential spectrum is tuned to the fundamental (pumping) frequency $f_{\beta}$ (the low-frequency coherent structures) in this case. The practical smooth predictability takes place at least up to the fundamental period $T = 1/f_{\beta}$. Generally, the fundamental (pumping) periods indicated in the Figs. 3-14 can be considered as practical limits for the smooth predictability of the surface air temperature dynamics for these cases. It means that the practical smooth predictability of the air surface temperature dynamics is possible {\it at least} up to this time scales.

\section{Conclusions}

 A. The apparent universality of the decaying part of the surface air temperature spectra on the daily to intraseasonal time scales can be related to the Hamiltonian distributed chaos. Actually just the appropriately (accordingly to the Hamiltonian distributed chaos laws) chosen scales on the figures 3-14 allowed to reveal this universality. Of course, these are examples only and there should be a comprehensive analysis of as numerous as possible local records of the surface air temperature in order to check out this universality and to find out its limitations.\\
  
  B.  For Central England there is no significant difference in the universal spectrum between a Modern (1944-2018yy) and an Historical (1858-1943yy) periods (cf. Figs 5 and 6). \\
  
  C.  It seems that the value $\beta =3/4$ corresponds to the most of the mid-latitudes, whereas the value $\beta =1/2$ corresponds to the high-latitudes and to tropics.  \\
  
 D. The fundamental (pumping) period $T_f \simeq 40$ day seems to be a universal one for the Northern Hemisphere. The near 40 day oscillations at the Northern Hemisphere extratropics are well known from numerous observations and are theoretically associated with the interaction of topography with the non-zonal flows (cf. Refs. \cite{mag}-\cite{cun} and references therein).\\ 
 
 E. The global average surface temperature fluctuations (land-based) are also dominated by the Hamiltonian distributed chaos with $\beta =1/2$. The chaos is sensitive to a half-year period (the $T_0$ is approximately equal to this period in the Figs. 16 and 17). The shift of the fundamental (pumping) period $T_f$ toward 50 days can be related to the effect of the Madden-Julian oscillation (see, for instance, Refs. \cite{wh},\cite{ven2},\cite{b3},\cite{mj}-\cite{hms} and references therein) but it can be also related to the large-scale space averaging (cf. Figs. 5 and 6). \\
 
 F. The tropical Pacific SST fluctuations on the daily to intraseasonal time scales are dominated by the Hamiltonian distributed chaos with $\beta =1/2$ (Fig. 20). The fundamental (pumping) period $T_f \simeq 57$d can be compared to the corresponding periods of the Madden-Julian oscillation (cf. Refs. \cite{wh},\cite{ven2},\cite{b3},\cite{mj}-\cite{hms}). It is not surprising because the Madden-Julian oscillation is a dynamic coupling between tropical ocean deep convection and large-scale atmospheric circulation. The second peak in the spectrum shown in Fig. 20 corresponds to period $T\simeq 33$d and can be compared to the fundamental (pumping) periods characteristic to the surface air temperature in the Australia and South America (cf. Figs. 11, 13, 14 and 19). \\
 
G. The practical smooth predictability for the surface air temperature dynamics is possible at least up to the fundamental (pumping) period of the distributed chaos.

\section{Acknowledgement}
I acknowledge use of the data provided by the Weather, climate and hazards service of the Government of Canada, the Aisori Meteo Servise (Russia), the Met Office Hadley Centre observations datasets (UK), the European Climate Assesment\&Dataset, the KNMI Climate Explorer, the Berkeley Earth dataset (USA) and the National Centers for Environmental Information, NOAA (USA). I thank Y.Q. Sun and F. Zhang for sharing their data and discussions.

\end{document}